\documentclass{PoS}
\title{Black Holes as Cosmic Dynamos}
\ShortTitle{Black Hole Dynamos}
\author{\speaker{Roger Blandford}
\\KIPAC, Stanford University, USA\\E-mail: \email{rdb3@stanford.edu}}
\abstract{An introduction is given to a meeting on the role of massive and stellar black holes in powering non-thermal activity in a rich  variety of cosmic sources. Relevant properties of magnetized, spinning black holes are summarized and their observational expression, within galactic nuclei, in terms of radio loudness and Fanaroff-Riley class, is briefly described. The dependence of the accretion mode on the rate and manner of the mass supply beyond the black hole sphere of influence is also discussed. It is argued that hydromagnetic outflows from accretion disks are generally expected over as many as six decades of radius and that they may be the source of emission line gas. These outflows collimate the relativistic jets which are probably generated in an electromagnetic form but become hydromagnetic as they entrain gas through boundary layers where most of the initial nonthermal emission occurs. It is proposed that the particle acceleration close to the hole emphasizes the proton channel which allows secondary pairs to be created at far higher energies than is possible from direct acceleration. These pairs radiate synchrotron $\gamma$-rays which can escape along the jet because the outflow effectively shields them from pair-producing, soft photons. Jets are subject to helical instabilities which can tangle their magnetic field and may destroy them. The jet should become plasma-dominated through intermittent, ``magnetoluminescent'' untangling of the field which causes nonthermal emission all along its length. Powerful jets remain supersonic out to the ``hot spots'' at the extremities of the source; weaker jets become subsonic plumes or bubbles. The prospects for learning much more about the nature and operation of jets over the next decade are excellent.}
\FullConference{International Conference on Black Holes as Cosmic Batteries: UHECRs and Multimessenger Astronomy - BHCB2018\\12-15 September, 2018\\Foz do Igua\c cu, Brasil}
\begin{document}
\section{Introduction}
We have just celebrated the centenary of Curtis' \cite{curtis18} discovery of the M~87 jet which, with the benefit of hindsight, came a decade after the discovery of the first Active Galactic Nucleus or AGN, NGC 1068 \cite{fath09}, at the same (Lick) observatory. We now know AGN are powered by spinning black holes and the gas that they accrete, as well as stellar processes \cite{meier12}. In addition, stellar black holes are observed as powerful non-thermal emitters and relativistic jet creators in X-ray binaries \cite{mirabel99} and Gamma Ray Bursts \cite{kouveliotou12}. (Analogous processes are probably involved in the formation of jets by neutron stars \cite{durant13} and protostars \cite{bally16}.)

This introduction will focus on one particular interpretation of relativistic jets, namely that their power is extracted from the spin energy of the black hole through the agency of strong electromagnetic field \cite{blandford18}. The evidence in favor of this view has strengthened in recent years but alternatives where the agency is essentially fluid and/or the source is orbiting gas can still be entertained. However, they will not be rehearsed here. 
\section{Spinning Black Holes}
\subsection{Theoretical Black Holes}
It is now 235 years since John Michell considered a body so compact that light would not have the escape velocity and 103 years since Schwarzschild found what we now know to be the general relativistic solution for a non-spinning, spherically symmetric black hole. This solution was, quite remarkably, generalized by Kerr \cite{kerr63} to a spinning hole.  After several decades of confusion, research teams, led by Sciama, Wheeler and Zel'dovich essentially laid down the modern, physical interpretation of a black hole based on the application of novel mathematical methods by Penrose, Hawking and many others.

The black holes observed by astronomers are essentially very simple (although more general black holes have led to important insights into theoretical physics). They are described by their mass, $M$, (as measured by a distant satellite) expressed as a gravitational radius $m\equiv GM/c^2=(M/10^8\,{\rm M}_\odot)\,{\rm AU}$ and their angular velocity $\Omega_{\rm H}<1/2m=10^{-3}(M/10^8{\rm M}_\odot)^{-1}\,{\rm rad\,s}^{-1}$ which is all that is needed to describe the geometry. As accreted mass typically has more specific angular momentum than a hole, it is reasonable to suppose that most holes are spinning rapidly unless the direction of the accreted angular momentum keeps changing \cite{hughes03}.  The black hole is surrounded by an event horizon --- a cosmic Igua\c cu Falls --- of radius $r_{\rm H}=2m/(1+4\Omega_{\rm H}^2m^2)$. The area $A=4\pi r_{\rm H}|^2$ of the event horizon can be shown to be non-decreasing, like the entropy to which it is proportional. This allows us to define an irreducible mass $m_0=(A/16\pi)^{1/2}=m(1+4\Omega_{\rm H}^2m^2)^{-1/2}$ and a spin energy $m-m_0$ which can be as large as $0.29m$ and which is, in principle, entirely extractable. 

The simplest way to demonstrate extraction \cite{penrose69} is to imagine a small mass, $\mu$, free-falling from rest at large distance, where its energy is $\mu c^2$, towards the black hole. It turns out that close to the horizon, there are orbits with negative total energy, including rest mass. Now, suppose that the mass splits into two pieces in a manner that conserves 4-momentum and one piece is placed on a negative energy orbit that crosses the event horizon, reducing the mass of the black hole, though increasing the irreducible mass. As energy is conserved, the other piece must escape with finite kinetic energy at infinity. Rotational energy is extracted in principle, though this mechanism is unlikely to be important in practice.

A more realizable method of extracting the spin energy uses electromagnetic field supported by external current and charge \cite{blandford77}. We replace particle orbits with a solution --- either perturbative or numerical --- of the Einstein-Maxwell equations in the Kerr spacetime, subject to the requirement that the field be finite as measured by an observer crossing the event horizon and satisfy astrophysical boundary conditions at large distance. These equations can either be supplemented with the force-free prescription --- the divergence of the electromagnetic stress-energy tensor vanishes --- or a relativistic magnetohydrodynamical, MHD, description. Electromagnetic energy will be transported as a Poynting energy flux and can flow out of the hole in a non-physical frame that is non-rotating with respect to flat space at infinity. However when we transform into the rest frame of an observer hovering just outside the horizon with angular velocity $\Omega_{\rm H}$ the energy flow will be inward, as expected. Relativistic spin/gravitational energy is no more localizable than the position of an electron in phase space and it is not possible to point to a specific source of the extracted energy; it suffices to demonstrate global conservation of 4-momentum.

\subsection{Observed Black Holes}
\subsubsection{Galactic Black Holes}
The first black hole to be convincingly identified is Cyg-X-1 \cite{webster72}, where the compact object is now known to have a mass $\sim15\,{\rm M}_\odot$, over five times the maximum credible mass for a neutron star. Roughly twenty of these X-ray binaries are well-studied and a minority have prominent, relativistic jets. The best studied Galactic jet source is SS 433, which produces two anti-parallel jets,with speed $\sim c/4$. These precess on a cone with angle $\sim20^{\rm o}$ every $\sim162\,{\rm d}$. The black hole mass is estimated to be $\sim20\,{\rm M}_\odot$, while the companion ``A'' star could actually be a binary as this would provide an explanation for the precession. Interestingly, SS 433 has just been detected as a $\gamma$-ray source with photons up to $\sim25\,{\rm TeV}$ \cite{abeysekara18}.  

\subsubsection{Gamma Ray Bursts}
Gamma Ray Bursts (GRB) are now generally regarded as heralding the birth of a stellar black hole either in a core collapse supernova or in a binary neutron star merger. Both classes are associated with relativistic jets, which are generally attributed to electromagnetic effects around magnetized spinning black holes \cite{kouveliotou12}.

\subsubsection{Massive Black Holes}
Most normal galaxies appear to have compact nuclei containing central black holes with masses in the range $\sim 10^5-10^{10}\, {\rm M}_\odot$ \cite{meier12}. Some of these are active and have luminosity that can be as large as $\gtrsim10^{48}\,{\rm erg\,s}^{-1}$ in the case of the very brightest quasars. They can also be relatively dim, a good example being our $4\times10^6\,{\rm M}_\odot$ black hole which has a bolometric luminosity of $\sim10^{36}\,{\rm erg\,s}^{-1}$ \cite{abuter18}.  

The history is instructive. NGC~1068 is now known as a Seyfert galaxy. These are generally associated with spiral galaxies and are radio-quiet (though not silent) and the luminosity is presumably dominated by the accretion disk. By contrast, the jet power in M87 appears to exceed the disk luminosity by at least several hundred. M~87 is a ``Fanaroff-Riley'' I, FR~I, source \cite{fanaroff74} as the jet is still insufficiently powerful to remain supersonic beyond the galaxy. The first powerful radio galaxy was Cygnus A which was shown to comprise two radio lobes outside the galaxy that we now know to be powered by FR~II jets that remain supersonic to the outer reaches of the lobes where they terminate in ``hot spots''. The first quasar, 3C~273 \cite{gravity18}, is radio loud and one of its jets is pointed towards us so that it exhibits superluminal motion suggesting that the jet speed has a physical Lorentz factor $\Gamma\sim10$.

\subsection{Multi-Messages}
\subsubsection{Gravitational Radiation}
The study of black holes has been invigorated by observations and connections to the three additional ``messengers''. The discovery of ten black hole mergers by LIGO-VIRGO has extended the upper limit for ``stellar'' black holes to $80\,{\rm M}_\odot$, while the single neutron star merger almost certainly shows the formation of a black hole with mass as low as $2.8\,{\rm M}_\odot$. This source, GW~170817, also produced a pair of relativistic jets associated with a low power, off axis GRB \cite{mooley18}.

\subsubsection{Neutrinos}
There is now a tentative identification of a single $\sim290\,{\rm TeV}$ neutrino with the blazar TXS~0506+056 \cite{aartsen18} and we could be on the verge of opening up this important channel in jet observations. 

\subsubsection{Cosmic Rays}
AGN and GRB jets have long been considered a prime candidate for the source of cosmic rays with energy in the $\sim\,{\rm PeV}$ to $\sim\,{\rm ZeV}$ range \cite{hillas84}. As suggested above there is enough EMF to account for the Ultra High Energy particles. The problem is that the high radiation energy density in these sources may preclude their escape.

\section{Disk and their Outflows}
\subsection{Accretion Flows}
The black hole dominates the gravitational potential within a radius of influence, $r_{\rm inf}\equiv GM/\sigma^2$, where $\sigma$ is the ID central velocity dispersion, which is similar to the Bondi radius. Within this radius, gas is thought to have sufficient angular momentum to form an accretion disk. The behavior of the disk is generally supposed to depend heavily upon the ratio $\dot m$ of the mass supply $\dot M$ relative to the critical rate, $\dot M_{\rm crit}=4\pi GMm_{\rm p}/\sigma_{\rm T}c$ \cite{meier12}. However, we do not have a well-accepted description of accretion onto compact objects, only competing theories that are all challenged by some observations. Simulations, which have improved our understanding of fundamental principles, cannot yet make a confident connection to observation through a detailed understanding of microphysical processes like radiative transfer, particle acceleration, wave-particle plasma interactions and dynamo action.

\subsubsection{Intermediate Mass Supply}
``Classical'' accretion disk theory may be applicable when $\dot m\sim1$. In this case, the gas can radiate efficiently and cool at all radii and so the disk remains thin. Ionized disks are subject to the magnetorotational instability \cite{balbus98}, which causes the magnetic field inside the disk to grow in a few orbital periods. The nonlinear steady state has been extensively simulated and is now being studied at the plasma level. Consider a thin, stationary disk. The magnetic torque $G(r)\propto r^{1/2}$ is responsible for the outward flow of angular momentum and does work at a rate $G\omega$ where $\omega$ is the Keplerian angular frequency. This is the energy flux and its divergence --- twice the rate of binding energy release by the infalling gas also has to be radiated and so the total flux radiated is three times the rate of release of binding energy. 

\subsubsection{Low Mass Supply}
When $\dot m<<1$, the flow may also be unable to radiate efficiently. Here, the argument is that the Coulomb scattering, electron-ion equilibration time can be longer than the inflow time \cite{yuan14}. If we ignore the very real possibility that plasma wave-particle interactions effect a faster coupling, then the ions near the hole will heat to $\sim100\,{\rm MeV}$ temperature while the electrons remain at $\sim\,{\rm Mev}$ energies. Three outcomes have been discussed. (i) The ions form a thick disk or torus and there is conservative flow onto the hole with low radiative efficiency. (ii) The liberated energy is high and escapes from close to the hole through convection or bubbles. (iii) Most of the liberated binding energy is diverted into an outflow starting at large radius. Option (iii) seems to be inevitable because the surplus energy transported by the torque must be removed \cite{blandford99}. Thickening the disk makes the arithmetic model-dependent but does not obviate non-radiative outflow. Sometimes this is called ``altruistic accretion'' because a few protons sacrifice themselves so that the majority of their fellow ions can escape to freedom!

The external torque acting on the disk surface \cite{blandford82} depends upon the magnetic fieldlines passing through it. However, there is a crucial dynamical difference. A hydromagnetic surface stress need not incur any dissipation unlike the magnetic stress acting inside the disk. In principle, it can remove energy and angular momentum in just the right ratio to allow the gas to flow inward without entropy production. Unlike what happens with the event horizon, there need be little effective surface resistivity. In the limit, a strong, external torque can cause cold, invisible accretion.

Real disks are, assuredly, more complex. Closed flux tubes that connect one radius with another will be stretched by the differential rotation and forced to reconnect above the disk in a corona. Some of the heat will appear as particle acceleration; some will be transported away as hydromagnetic waves. Even if the flux tubes are open, they are likely to form individual, current-carrying, flux ``ropes'' twisted around the rotation axis in the same sense by the differentially rotating disk.

\subsubsection{High Mass Supply}
When $\dot m\gtrsim10$, the inflow is large enough to produce a luminosity above the Eddington limit if the radiative efficiency is high. However, Thomson scattering traps the photons and there are the same three options as when $\dot m<<1$. Again the evidence favors mass loss at all radii driven by radiation pressure and magnetic stress. The gas in the outflow will transition from optically thick to thin as its density decreases. Large final speeds are possible if the resonance lines of partially ionized metals dominate the opacity as may happen in Broad Absorption Line Quasars. Dust can be important at large radii. The rate of mass loss is very hard to estimate, just as is the case with the sun.

\subsection{Jet Collimation}
mm VLBI \cite{kim18,giovannini18} and $\gamma$-ray variation \cite{aharonian07,ackermann16} demonstrate that jets are accelerated and collimated at small radius \cite{blandford18}. They also exhibit strong limb-brightening which strongly suggests the formation of a boundary layer. Many simulations have presumed that jet collimation is essentially gas dynamical and due to the funnel created by an extensive, slowly inflowing torus \cite{nakamura18}. This is problematic because jets are still being collimated at many thousands of $m$, and there is little evidence for any disk when the mass supply is low.. A further concern is that electromagnetic jets are subject to helical instability \cite{hardee99} which might easily disrupt them when confined in this fashion.

An alternative view is that the disk is always thin or inconsequential and is strongly magnetized over many decades of radius and it is the MHD outflow which is directly responsible for jet collimation from the horizon outward. If the magnetic pressure exceeds the gas pressure in the disk, the Alfv\'en speed will also exceed the sound speed and the magnetorotational instability will be suppressed. (The Alfv\'en speed must be less than the circular velocity, though.) The vertical field strength should roughly match the hole field at the inner edge of the disk and the angular frequency will likely be $\sim\Omega_{\rm H}/4$ to match that of the field lines threading the equator of the hole. At larger altitude, where the jet becomes relativistic and the magnetic field, becomes toroidal, outward magnetic pressure  will be partially offset by electric tension in the jet and the ``hoop''stress of the sub-relativistic disk outflow can collimate the jet.   

The MHD outflow, must also be collimated. It is envisaged that nested magnetic surfaces extend over several decades of cylindrical radius. The poloidal field at larger radius confines the toroidal field at smaller radius. This continues until the accreting gas is encountered, perhaps at $r_{\rm inf}$. Suppose that there is a certain total magnetic flux confined within this outer radius. A uniform magnetic field threading the disk would neither create a jet nor affect the dynamics of the inner disk. Instead, the field must be centrally concentrated for the jet to be a large current source. A negligible fraction of the total flux suffices to achieve this, however. If flux concentration happens, then a quasi-stable electro/hydromagnetic configuration may be set up with $B\propto R^{-k}$. If stress balance in the outflow, rather than the inertia of the disk that is important, then arguments can be given for $k$ to lie in the range 1 to $5/4$. In this case, a further, simple argument gives a rough estimate for the jet power $\sim \dot M\sigma^{2-k}c^k$. Something similar may happen in pulsar wind nebulae, which can also exhibit axial jets \cite{durant13}.

Flows of this general character are prone to non-axisymmetric instability and it is entirely possible that relativistic jets confined by MHD outflows are only sufficiently stable to escape the black hole sphere of influence when the total magnetic flux that is trapped is large enough to suppress the magnetorotational instability. This, in turn, may require that the accretion onto the galactic nucleus be quasi-spherical, as should happen in elliptical galaxies. Accretion through a disk may not trap flux this effectively, thereby providing the basis for an explanation for the difference between radio-quiet and radio-loud AGN. Failed jets, associated with the former class, may still produce magnetic flux tubes which become ``wrapped around the axle''. The associated dissipation provides a plausible site for the ``lamppost'' inferred to illuminate X-ray Seyfert galaxy disks \cite{yuan19}.          

\subsection{Mass Loss}
\subsubsection{Centrifugal Launching}
There is also likely to be a significant mass flux in the outflow. Imagine a rigid fieldline attached to the disk. Protons and electrons, with small Larmor radii, will behave like beads on wires and be flung outward (or inward) if the magnetic field is inclined with respect to the vertical by more than $30^\circ$.  The details of how plasma is launched on this journey at the disk surface are unclear and the discharge is quite possibly time-dependent. In a MHD wind, the field lines will be isorotational with the disk at the radius of the footpoint and the outflow should transition to a supersonic state after passing through slow, intermediate and fast critical points. As the ``lever arm'' can be much larger than the radius of the footpoint of the flux tube, the torque acting on the disk and consequently the luminosity of the outflow can be quite large. When this is the case the mass lost by the accreting disk will be small. However, it is also possible that significant mass is lost from all radii through this mechanism and, consequently, the flow across the event horizon is a tiny fraction of the mass supply at large radius.  

\subsubsection{Broad Emission Line Clouds}
A defining feature of quasars and Seyfert galaxies is the presence of broad emission lines. These are formed by compact regions of $10^4\,{\rm K}$ gas moving with speeds $\sim10,000\,{\rm km\,s}^{-1}$ at a distance $\sim0.1\,{\rm pc}$, filling factors $\sim10^{-5}$ and covering factors $\sim 0.1$. The provenance, dynamics and fate of this gas is a major puzzle. One possibility is that it is the gas flung out from the disk \cite{bottorff97}. The magnetic field solves the problem of cloud confinement but it is not clear if this rather specific kinematics is consistent with line profiles, reverberation studies and infrared interferometry \cite{gravity18}. The quantity of gas required, typically a few solar masses, is quite small.

\subsubsection{Jet Shielding}
A closely related possibility is that the gas leaving the innermost disk, which should be a minor contributor to the total emission line profile, may provides an effective shield of the inner jet, efficiently absorbing photons from the Lyman edge to the X-ray band. This would then allow GeV $\gamma$-rays, which can vary on timescales as short as minutes in quasars, to avoid pair production and escape along the jet. The flow of mass required to do this can be quite small. It would be necessary for the gas to undergo sufficient expansion cooling for the hydrogen to recombine \cite{konigl94}. 

\section{Relativistic Jets}
\subsection{Electrodynamics}
Now, let us turn to the jets themselves. If we assume that the electromagnetic field is stationary, axisymmetric and force-free, we can replace the horizon with an equivalent Newtonian surface within which surface current density $J_{\rm H}(\theta_{\rm H})$ flows instead of crossing the horizon, with $\theta_{\rm H}$ being the polar angle at the horizon. (This device is not especially helpful when there is time-dependence.) The boundary condition allows us to equate $J_{\rm H}$ to the potential gradient times the impedance of free space $Z_0=\mu_0c=377\,{\rm Ohm}$. Now, the assumptions that we have made imply that the angular velocity, $\Omega$, of a magnetic field line or, more precisely, that of an observer for whom the electric field vanishes, is constant along field lines along with the electric potential. The conserved energy and angular momentum fluxes and current also flow, poloidally, along the field lines. 

If the magnetic flux threading the hole is $\Phi(\theta)$, and $V(\theta)$ is the potential, then the Maxwell equations imply $dV/d\Phi=\Omega/2\pi$. In addition, a continuous current flows from the horizon to large radius while the charge carriers --- electrons and positrons --- flow inward at the horizon and outward in the far field. This requires sufficient pair production within the magnetosphere to sustain the current. Given the enormous potential, this is not hard to imagine happening, though several competing mechanisms have been discussed \cite{chen18}. There are obvious parallels with models of radio pulsar magnetospheres.

$\Phi(\theta)$ is determined by imposing stress balance in the magnetosphere and at its boundary. If the outflow terminates in a relativistic jet, there will also be a radiation condition that $E\sim cB$. In general, we can use a circuit description to characterize solutions in terms of a battery producing a voltage $V$, a current $I$ flowing through the hole and through an effective ``load'' at many gravitational radii \cite{thorne88}. Typical solutions have $\Omega\sim0.4\Omega_{\rm H}$, while the effective impedances of the hole and the load are $\sim50\,{\rm Ohm}$ so that $V({\rm V})\sim100I\,({\rm A})\sim(50L_{\rm jet}({\rm W}))^{1/2}$ and power is ``dissipated'' in both the hole (increasing $m_0$) and the load (producing relativistic particles and nonthermal radiation). $V$ ranges from $\sim10\,{\rm PV}$ for weaker Galactic superluminal sources to $\sim100\,{\rm ZV}$ for the most powerful Gamma Ray Bursts.

In the case of a force-free solution, the velocity is only defined perpendicular to the magnetic field. However, if we take the poloidal component then this will increase, roughly linearly with radius. However, linear momentum will also be transported outward across the boundary layer to decelerate the flow, while ionic plasma will be entrained into the jet \cite{hardee99} to increase the proton content and to effect a transformation to a relativistic MHD flow where a velocity can be defined by the center of momentum frame of the plasma. Given this complex velocity field, it is quite likely that the emission seen by a distant observer will be dominated by those regions where the jet velocity is inclined to the line of sight by an angle $\sim\Gamma^{-1}$. 

\subsection{Magnetohydrodynamics}
A MHD jet will probably reach an equilibrium $\Gamma$ where the acceleration associated with the expansion will be balanced by deceleration due to surface interaction and instability. It will likely remain relativistic until the gravitational potential variation changes around $r_{\rm inf}$ which is where jets are most vulnerable and their profiles and dissipation change. High power jets seem to make it through $r_{\rm inf}$ and remain at least mildly relativistic and supersonic until they reach the ``hot spots'' at the extremities of the radio lobes --- the FR~II sources. Weaker jets fall prey to the abrupt ``breaks'' in the external medium which create recollimation shocks around $r_{\rm inf}$ and convert to subsonic, buoyant plumes and, eventually, bubbles which float away from the galaxy --- the FR~I sources.

\subsection{Particle Acceleration}
\subsubsection{Diffusive Shock Acceleration}
Shocks are weak and ineffective accelerators when the jet flow is magnetically-dominated. However, outer jets in FR II sources appear to become plasma-dominated and the hot spots are presumably strong shocks where the relativistic electrons are accelerated and radiate \cite{matthews18}.

\subsubsection{Relativistic Reconnection}
Most recent attention has been devoted to relativistic reconnection \cite{beloborodov17} which must occur within the magnetically-dominated inner jets. A typical starting 2D configuration is a current sheet which can break up into a one dimensional array of X-points separated by magnetic islands. The current becomes so strong close to the X-points that significant resistance develops leading to ohmic dissipation and heating. (The resistivity should be treated as a tensor and the Hall effect can be very important.) Magnetic field lines exchange partners within this small reconnecting volume. Non-relativistic reconnection is generally rather inefficient for particle acceleration with much of the released magnetic energy being taken up by the thermal plasma and MHD waves. 

Impressive, Particle In Cell Simulations have been performed for relativistic reconnection and these demonstrate high efficiency. Energetic tails to the particle distribution function are produced and much of the magnetic power density can emerge in relativistic electrons and, when present, positrons. The actual electrodynamic mechanisms involved are varied involving drift motion along large electric fields, adiabatic compression inside collapsing islands and second order Fermi acceleration by moving magnetic structures. Three dimensions, of course, complicates matters. Firstly there is the influence of a slowly varying ``guide'' field which is simply convected by the moving plasma and can seriously change the individual particle orbits. Additional structure along the third dimension can expand the topological complexity and create more reconnection sites.

Despite its likely importance in steady magnetic dissipation, relativistic reconnection seems to be poorly suited to accounting for the rapid variability of $\gamma$-ray sources which require the conversion of electromagnetic energy into $\gtrsim\,{\rm GeV}$ particles over large volumes at the speed of light. Reconnection is intrinsically slow as large amounts of magnetic flux has to pass through a small area.  

\subsubsection{Magnetoluminescence}
If relativistic jets are as strongly magnetized, as we assume, then the Larmor radii of all charged particles, except, possibly, Ultra High Energy Cosmic Rays, are tiny in comparison with the jet. Their drift motions are extremely slow and are responsible for the electrical current component perpendicular to the local magnetic field \cite{thorne17}. Under conditions of ideal MHD or force-free electrodynamics, ${\bf E}\cdot{\bf B}=0$ and the component of current density along the field is $({\bf B}\cdot\nabla\times{\bf B}/\mu_0-\epsilon_0{\bf E}\cdot\nabla\times{\bf E}){\bf B}/B^2$.

A relativistic magnetic rearrangement may come about if the magnetic field is organized into a network of space-filling flux ropes, rather like those observed, for example, in the solar corona and the Galactic center. The ropes will carry volumetric current along them and much higher current density at their interfaces. (Just like real ropes these ropes will be twisted, due to the presence of the current, and can be composed of thinner strands and have ``hairy'' surfaces.) Instabilities and velocity gradients in the inner jet can cause these ropes to become tangled. As the jet expands, the tension along the ropes will cause them to untangle without topological change. This can happen quite abruptly. In the absence of dissipation the electromagnetic energy released will be carried off by hydromagnetic wave modes. 

However, when one rope slides past another and the current sheet becomes so thin that there are insufficent charges to carry the current density, a substantial component of ${\bf E}\cdot{\bf B}$ will develop and the electric shear stress will contribute a frictional force that will lead to dissipation in the current sheet. The speed with which these flux ropes untangle will increase until the tension balances the friction. Furthermore if the plasma is ionic, then half the dissipated energy will go into accelerated protons. (Note that this process is distinct from reconnection, where the relative velocity of the magnetic field zones is perpendicular to the current sheet not parallel.)  We call this untangling, and the subsequent radiation, ``magnetoluminescence'' \cite{blandford17}.

\subsection{Nonthermal Emission}
\subsubsection{Inverse Compton Scattering}
Inverse Compton scattering is commonly invoked to account for the high energy hump in the ``Bactrian'' spectrum characteristic of blazars \cite{madejski16}. However, this is very hard to reconcile with the idea that the emission site is magnetically dominated. This is because the high energy hump is commonly higher flux than the low frequency hump. This implies that the magnetic energy density is smaller than the radiation energy density which, in turn, is smaller than the particle energy density presuming the same particles are responsible for both humps. This ``one zone'' model is quite implausible given that emission is observed over ten decades of radius in many jets --- consider a one zone model of the sun as a $6000\,{\rm K}$ black body and wondering how the neutrinos are made --- and multizone models have therefore been developed. However, these are generally incompatible with our assumption of electromagnetically-dominated inner jets unless the emission comes from very large radius which leaves the challenge of rapid variability.

\subsubsection{Synchrotron Radiation}
The low energy spectral hump is commonly associated with synchrotron radiation although some high apparent brightness temperature, variable radio emission has been associated with cyclotron masers and other coherent processes. If the radio emission is synchrotron radiation, then the ``core'' observed with VLBI is really a photosphere where the optical depth to synchrotron self-absorption is unity. This photosphere receded towards the hole as the frequency increases, observed effect known as ``core shift''. 

A more radical idea is that the high energy hump is also electron synchrotron radiation \cite{meyer18}. An immediate objection is that if the accelerating electric field satisfies $E\lesssim cB$, then the minimum emitted wavelength is the classical electron radius, $\equiv70\,{\rm MeV}$. However, if protons are accelerated in the jet boundary layer or a current sheet separating different magnetic ropes, they can be accelerated by the electric field to  PeV-EeV energies until radiation reaction due to Bethe-Heitler photo-pair production, with cross section $\sim\alpha\sigma_{\rm T}$, sets in. If, for example, there is jet shielding above $\sim10\,{\rm eV}$ and the effective jet Lorentz factor is $\Gamma\sim10$, then PeV protons will produce $\sim50\,{\rm MeV}$ pairs in the proton rest frame. These will be boosted to $\sim50\,{\rm TeV}$ in the jet comoving frame and they can cool rapidly, by emitting $\gamma$-ray synchroton radiation up to $\sim10\,{\rm GeV}$ in the black hole frame. So long as jet shielding is effective, these photons can avoid pair-production and escape the AGN. Note that this mechanism is intrinsically highly efficient because essentially half the electromagnetic energy finds its way into gamma rays so long as the accelerating electric field is strong enough to be balanced by radiation reaction. BL Lac objects lack these soft photons and emit $\sim{\rm TeV}$ $\gamma$-rays which also vary on minute timescales. They do not need to be shielded from disk photons to escape but the soft, nonthermal jet spectrum must originate at much larger radius for this to be possible.

\subsubsection{Pion Production}
Photopion production is also possible when the photon energy in the proton rest frame exceeds threshold,  $\sim150\,{\rm MeV}$ \cite{muecke03}. This can obviously make High Energy neutrinos as may have been observed, but the neutrinos and neutrons can escape the source and can lead to low radiative efficiency.

\section{Current Challenges and Future Observations}
While our understanding of the theory of the magnetized black holes, the topic of this meeting, is now relatively secure, our handling of the dynamical and radiative properties of disks and jets is not. Even basic matters, such as the location of the sources of the various spectral components of the disk and the jet, are controversial. Simple kinematical prescriptions lack a dynamical context and the rules for particle acceleration --- the crucial link between numerical simulations and observations of jets --- is highly conjectural. The main factors dictating AGN taxonomy are, likewise, disputed.

However the prospects for imminent change are good. The most immediate new capability is the Event Horizon Telescope \cite{doeleman12} which is making mm VLBI observations of Sgr~A$^\ast$, M~87 and other sources, anchored by ALMA which should start to resolve these sources on scales of a few gravitational radii. In particular, it will be of great interest to see if the mm source in Sgr A* is consistent with a thick ion torus, as commonly presumed, or if it indicates the presence of a very weak, and possibly failing, jet, as suggested here. The recent observations using the GRAVITY instrument on the VLT are providing complementary information. JWST and the thirty meter class optical telescopes, currently under construction, will resolve the inner parts of galaxies and should help us understand the nature of the gas flow in the vicinity of $r_{\rm inf}$ and whether or not this is responsible for radio-loudness. In addition they will produce superior OIR images of jets on the small scale. The Cerenkov Telescope Array and the  water Cerenkov telescopes should solidify our understanding of the TeV variability and help us understand if $\gamma$-rays mostly originate from the inner jet as argued here or originate at much greater distance. X-ray polarimetry will be explored with IXPE while ATHENA should also resolve jets with far greater sensitivity and help pin down the extent of the X-ray source.

The three non-electromagnetic windows  --- the multi-messengers --- all look promising. It is only a matter of time before we see if $\sim{\rm PeV}$ neutrinos are commonly observed from blazars. The answer will have clear implications for jet models. The third LIGO-VIRGO observing run is expected to garner several neutron star mergers, presumably accompanied by jets, and should help us understand if, when and how, their primary working substance transitions from electromagnetic to hydromagnetic to gas dynamical. The contribution of AGN jets to the high energy cosmic ray spectrum will be better bounded by observation. 

In this talk, I have sketched some possible consequences of combining the hypothesis that cosmic batteries are the prime movers of relativistic jets with the increasingly challenging observations that are being made. The connection between these approaches will likely be mediated by new simulations where the boundary conditions reflect more what we know about conditions on the largest scales while accommodating our ignorance of much of the plasma microphysics and addressing the diversity of actual sources. However, most of all, the discovery space is so large and the observational capability are so promising that we can all look forward to even more radical interpretations than those tentatively advanced in this review.

\section*{Acknowledgements}
I am grateful to Rita de Cassia Dos Anjos and her colleagues for organizing a very stimulating and enjoyable meeting and her patience with a late contribution. I am indebted to my current collaborators including Richard Anantua, Mitch Begelman, Ke Fang, Noemie Globus, Dave Meier, Manuel Meyer, Tony Readhead, Jeff Scargle, Dan Wilkins and Yajie Yuan as well as past collaborators including Will East, Robert Emmering, Arieh K\" onigl, Amir Levinson, Kevin Lind, Greg Madejski, Jonathan McKinney, David Payne, Martin Rees, Isaac Shlosman, Sasha Tchekhovskoy, Roman Znajek and Jonathan Zrake for helping to formulate the perspective summarized here. Support by the Miller Institute and the Simons Foundation is gratefully acknowledged.

\end{document}